\documentclass[pra,preprint,amsfonts,amssymb,amsmath,titlepage,floatfix,showpacs,superscriptaddress]{revtex4} 

\usepackage{psfrag}
\usepackage{bm}

\usepackage{bm}
\usepackage{graphicx}  
\usepackage{latexsym}                
\usepackage{amsmath}     
\usepackage{amsfonts}  
\usepackage{amssymb}
\usepackage{amsthm}
\usepackage{dcolumn}
\usepackage{color}

\newcommand{\be}{\begin{equation}}
\newcommand{\ee}{\end{equation}}
\newcommand{\bea}{\begin{eqnarray}}
\newcommand{\eea}{\end{eqnarray}}

\def\softt{{\leavevmode\setbox1=\hbox{t}%
\hbox to \wd1{t\kern-0.6ex{\char039}\hss}}}

\begin{document}

\title{The role of the relative phase in the merging of two independent Bose-Einstein Condensates}

\date{\today}
\author{L.~F.~Buchmann}
\author{G.~M.~Nikolopoulos}
\affiliation{Institute of Electronic Structure and Laser, Foundation of Research and Technology Hellas, P.O. Box 1527, Herakleion 711 10, Crete, Greece}
\author{P.~Lambropoulos}
\affiliation{Institute of Electronic Structure and Laser, Foundation of Research and Technology Hellas, P.O. Box 1527, Herakleion 711 10, Crete, Greece}
\affiliation{Department of Physics, University of Crete, P.O. Box 2208, Herakleion 710 03, Crete, Greece}
\begin{abstract}
We study the merging of two independent Bose-Einstein condensates with arbitrary initial phase difference, 
in the framework of a one dimensional time-dependent Gross-Pitaevskii model. 
The role of the initial phase difference in the process is discussed, and various types of 
phase-sensitive excitations are identified.
\end{abstract}


\maketitle

\section{Introduction}
A large class of applications of Bose-Einstein condensation of dilute gases involves one way or another, the  
merging of two initially separate condensates.  Matter wave interferometry \cite{AtomInterf} and the quest for a 
continuous atom laser \cite{cwAL,chikkatur}, are only two prominent such applications.  
The process of merging requires the controlled unification 
of two trapped clouds. Prepared independently in separate traps, the two clouds are expected to be united 
into a single, trapped condensate of a well defined state.  One example of the successful merging of two independent 
condensates has been demonstrated  by Chikkatur {\em et al.} \cite{chikkatur}, through the use of optical tweezers 
to transport one 
condensate from their “production chamber” to merge it with a previously prepared one.  As expected, the final 
cloud was found to contain more atoms than either of the two initial clouds, but less than their sum; presumably due to evaporative 
losses.

Theoretical investigations and modeling of the process have so far been rather limited, with a number of relevant 
questions still open.  More precisely, in related work \cite{relatedwork, lewenstein, yi}, the influence of an initial phase 
difference between the two condensates on the dynamics of the merging, the question of the phase of the final 
condensate, as well as the type of phase-sensitive excitations were not addressed.  
These aspects are of vital interest as attested by their frequent recurrence in the relevant experimental literature \cite{chikkatur,Jo07}. 
Given that two independently formed condensates will have a random phase difference, it would seem that issues 
pertaining to the phase merit attention not only from a fundamental point of view, but also for the applications 
envisioned.  A case in point is highlighted by the recent experiments of Jo {\em et al.} \cite{Jo07} reporting 
the dependence of 
heating and atom loss during the merging on the phase difference of the fragments.
Motivated by the above theoretical and experimental developments, our aim in this paper is to explore somewhat 
further these aspects.  Although the work in this paper is limited to zero temperature, it does provide useful insight 
on the role of the initial relative phase on the merging.  

\section{The system}
\label{SecII}
The system under consideration pertains to two independent elongated condensates (L and R) 
consisting of a large number of bosonic atoms cooled into the lowest eigenmode of
the corresponding harmonic trap. The merging of the two condensates is achieved by bringing 
the two traps together in a controlled and adiabatic manner.  

\subsection{Model of the merging}
Following  \cite{lewenstein}, we have investigated this process 
in the context of a one-dimensional model. 
As we will see later on, despite its simplicity this model is capable of 
capturing many of the phenomena that take place during the merging process,  
and have not been addressed in earlier related theoretical work \cite{lewenstein,yi}. 

To be consistent with \cite{lewenstein,yi}, as well as the  experimental setup for condensate 
merging \cite{chikkatur}, we assume two nearly identical harmonic traps 
with confining frequency $\omega$. As the two traps move towards each other, the global potential
experienced by the trapped atoms can be modeled by a double-well potential of the form \cite{george,merzbacher}
\begin{equation}
V(x,t)=\frac{1}{2}\left(|x|-s(t)\cdot l\right)^2\label{sym},  
\end{equation} 
in dimensionless units. The function $s(t)$ determines the details of the merging 
(i.e., speed and time scale $T_{\rm m}$), which has to be 
adiabatic so that any kind of excitations due to movement of the traps are suppressed.
To this end, the transport of the condensates must take place on a
time scale much larger than the characteristic time-scale of excitations
along the merging direction, as well as the time-scale of interatomic 
interactions \cite{lewenstein,yi}. Moreover, excitations can be minimized 
by appropriately choosing the profile of $s(t)$.

Throughout our simulations, $s(t)$ has been chosen as 
\begin{equation}
s(t)=\left\{
\begin{array}{rl}
1 &\text{for } t =0, \\
\cos^2\left(\frac{\pi t}{2 T_{\rm m}}\right) &\textrm{for } 0 < t \leq T_{\rm m}, \\
0  & \textrm{for } t> T_{\rm m}. 
\end{array} \right.
\label{s_t}
\end{equation}  
In the beginning of the merging (i.e., at $t=0$) we have two well-separated traps, 
and the potential (\ref{sym}) exhibits minima at $x=\pm l$, with $l$ chosen sufficiently large. 
During the merging, i.e., for $0< t \leq T_{\rm m}$ the two traps approach each other, 
and the two minima of the double-well potential are located at $x=\pm s(t)l$.
Accordingly, the barrier between the two wells also decreases in this regime and in 
the end of the merging (i.e., at $t=T_{\rm m}$) we have complete overlap 
of the two condensates. After the merging, i.e., for $t>T_{\rm m}$, 
the potential remains a single harmonic well.

We choose to describe the evolution of the system by the time-dependent 
Gross-Pitaevskii equation (GPE), which in dimensionless units reads
\begin{eqnarray}
\text{i}\frac{\partial}{\partial 
t}\Psi(x,t)&=&-\frac{1}{2}\nabla^2\Psi(x,t)+V(x,t)\Psi(x,t)\nonumber \\
& &+g|\Psi(x,t)|^2\Psi(x,t),\label{tdgp}
\end{eqnarray} 
with the function $\Psi(x,t)$ normalized to unity. 
The nonlinearity parameter $g$ is proportional to the total number of atoms 
and the corresponding $s$-wave atomic scattering length. 
We use harmonic oscillator (h.o.) units, i.e. $\sqrt{\frac{\hbar}{m\omega}}$, $\omega^{-1}$ and $\hbar\omega$ 
for length, time and energy respectively, where $m$ is the atomic mass. 
The energy of a solution of Eq. (\ref{tdgp}) is given by
\begin{eqnarray}
E[\Psi]&=&\int\mathrm{d}x\left(\frac{1}{2}|\nabla\Psi(x,t)|^2+V(x,t)|\Psi(x,t)|^2\right.\nonumber\\
& &\left.+\frac{g}{2}|\Psi(x,t)|^4\right)\nonumber\\
&=&E_\text{kin}+E_\text{pot}+E_\text{int}\label{energy},
\end{eqnarray}
and is a conserved quantity provided that the external
potential does not depend on time \cite{review}. 
The lowest energy solutions of Eq. (\ref{tdgp}) to a given potential $V(x)$ can be written as
\begin{equation}
\Psi(x,t)=\psi(x)e^{-\text{i}\mu t/\hbar+\text{i}\varphi}.\label{gs}
\end{equation} 
Here, $\psi(x)$ is a real function whose squared modulus is the atomic 
density, while $\varphi$ is a random phase which emerges, as a broken gauge symmetry, 
in the process of creating a condensate  \cite{review}. 
The chemical potential $\mu$ is given by
\begin{equation}
\mu=(E_\text{kin}+E_\text{pot}+2E_\text{int})\label{chempot}.
\end{equation}
In general, the time-evolution of a condensate which is initially in its 
ground state, is uniquely determined by the two parameters $\mu$ and $\varphi$.

As mentioned earlier, in the beginning of the merging, the two identical condensates are considered 
to be  independent. This means that each condensate experiences its local harmonic potential only 
and thus, without loss of generality, the initial state of the system is given by     
\be
\Phi(x,t=0)=\Phi_{\rm L}(x)+\Phi_{\rm R}(x)e^{\text{i}\Delta\varphi_\text{in}},   
\label{gs_init2}
\ee  
with $\Phi_{\rm L(R)}(x)$ the lowest-energy solution for the isolated left(right) harmonic potential respectively, and $\Delta\varphi_\text{in}=\varphi_{\rm R}-\varphi_{\rm L}$ the initial phase difference between the two condensates, where $\varphi_{\rm R(L)}$ is the phase of the right (left) condensate.

\begin{figure}
\includegraphics[scale=1]{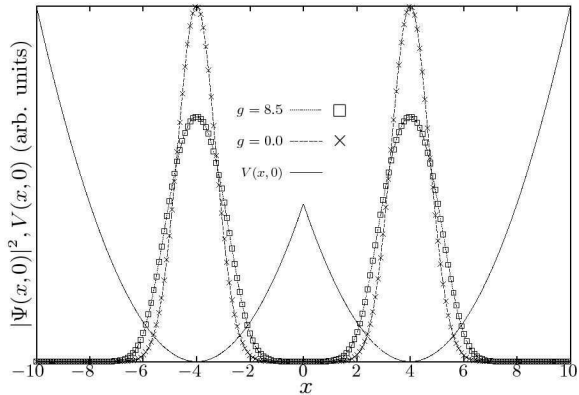}
\caption{
Initial conditions used in the numerical simulations. 
Symbols: atomic densities $|\Psi(x,0)|^2$ corresponding 
to the ground state of Eq. (\ref{tdgp}) for the double-well 
potential $V(x,0)$. 
Lines: atomic density distribution for two identical independent 
condensates trapped in harmonic potentials centered at $x=\pm 4$ (see Eq. \ref{gs_init2}). Harmonic oscillator units are used for $x$.} 
\label{groundstatedensity}
\end{figure}

In principle, however, this is not the case. According to Eqs. (\ref{sym}-\ref{s_t}) 
the two condensates at $t=0$ experience the double-well potential  $V(x,0)=(|x|-l)^2/2$. 
Hence, to ensure independence of the two condensates, throughout our simulations we had to choose 
sufficiently large separation $l$. In that case, the lowest-energy solution of the system 
$\Psi(x,0)$, is well approximated by Eq. (\ref{gs_init2}) (see Fig. \ref{groundstatedensity}), 
but the two condensates are always in phase, 
i.e., $\Delta\varphi_\text{in}=0$ . 
We had therefore to introduce by hand the initial phase difference between the two condensates, 
by multiplying $\Psi(x,0)$ by a factor $e^{\text{i}\Delta\varphi_{\rm in}}$, for $x>0$.  
Although the resulting state is no longer longer a ground state of the double-well potential, 
one may readily check that the increase in energy is very small due to the small 
value of $|\Psi(x,0)|^2$ at $x=0$. 
Finally, it is worth noting that the symmetry of the model for all $t\geq 0$ 
ensures that the initial phase difference, is not affected by the movement of the traps.  

\section{Numerical simulations}
\label{SecIII}
We solve Eq. (\ref{tdgp}) using a time-splitting spectral method \cite{numerics}, on a spatial 
grid of 2048 points ranging over 20 harmonic oscillator units. 
The size of the applied time step was $10^{-3}$ and the numerical performance -- estimated by the 
conservation of particle number and energy -- was found to be very good. Simulations were performed 
for various initial phase differences and merging times, in the absence, as well as in the 
presence, of interatomic interactions. 
In the latter case, the dimensionless nonlinearity parameter was chosen to be $g\simeq 8.5$, 
which characterizes an intermediate regime of interaction strength.
Throughout our simulations we have investigated the dynamics 
of the merging from the perspective of two quantities, namely 
the energy of the system and the phase of the final condensate.

It is worth keeping in mind for the following discussion that the system under 
investigation is invariant to changes of the global phase. 
This is obvious from the fact that the GPE which governs the evolution of the total wavefunction of the system 
remains invariant under transformations of the form $\Psi(x,t)\to e^{i\chi}\Psi(x,t)$, where 
$\chi$ is a constant global phase. Hence, such changes do not affect quantities that are determined only by 
densities $|\Psi(x,t)|^2$ and  phase differences (such as energy). Moreover, 
this invariance gives us some freedom in choosing a reference phase. Throughout 
our simulations, the reference phase has been chosen as $\varphi_{\rm L}=0$ [see Eq. (\ref{gs_init2})]. 

\subsection{Energy}
\label{IIIA}
\begin{figure}
\includegraphics[width=7.6cm]{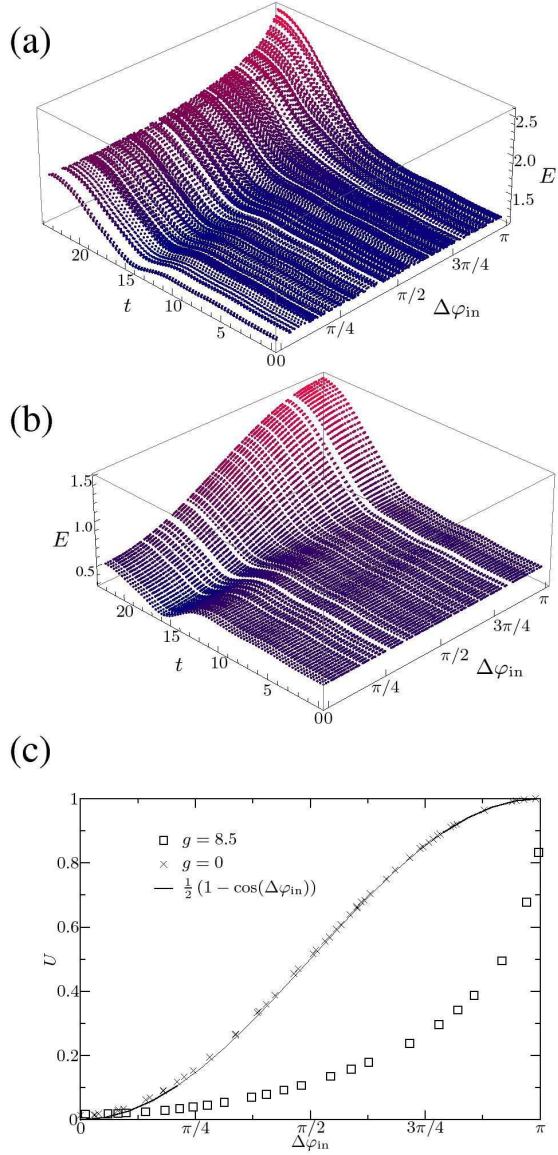}
\caption{(Color online)  
(a) Time-evolution of the total energy of an interacting gas during a merging process,  
for various initial phase differences.  (b) The same for an ideal gas.
(c) Excess energy at the end of the merging as a function of the initial phase 
difference $\Delta\varphi_\text{in}$. The dashed line through the 
crosses is the theoretical result for ideal gas, given by Eq. (\ref{nonintenergy}).  
For symmetry reasons, all the plots can be extended to $\Delta\varphi_{\rm in}<0$ by the 
transformation $\Delta\varphi_{\rm in}\to -\Delta\varphi_{\rm in}$. 
Merging times: $T_{\rm m}=22.5$ (a,b) and $T_{\rm m}=23$ (c).
Energy and time are in units of $\hbar\omega$ and $\omega^{-1}$, respectively.} 
\label{energylndscape}
\end{figure}

The time evolution of the energy of the system, as determined by Eqs. (\ref{tdgp}-\ref{energy}), 
for various initial phase differences is depicted in 
Figs. \ref{energylndscape}(a) and (b).  
At $t=0$ we find $E_{\rm id}(0)=0.5$ and $E_{\rm int}(0)=1.20$ for ideal $(g=0)$ and interacting 
gas $(g\simeq8.5)$, respectively. Clearly, due to the large initial separation of the traps, there 
is no remarkable dependence on the initial phase difference.  At the very early stage of the 
merging i.e., for $0<t\lesssim 10$, the overlap between the two condensates is negligible,  
and thus the energy remains very close to its initial value; a fact which also confirms the initial 
independence of the two condensates as well 
as the adiabatic nature of the merging process for the chosen parameters. 
For longer times (i.e., for $10\lesssim t<T_{\rm m}$), the condensates' wavefunctions begin overlapping 
in space, and the energy of the system depends crucially on the initial phase difference. 
In particular, as a general observation we note that the energy of the system 
increases as we increase the initial phase difference $\Delta\varphi_{\rm in}$.  
At $t\simeq 15$, we observe a local minimum in energy for small values of $\Delta\varphi_{\rm in}$.  
It occurs for both ideal and interacting gases; albeit at 
slightly different time instants. Hence, its appearance is not associated with the nonlinear 
character of the GPE, but is a feature of the particular double-well potential (\ref{sym}) 
and the energies of its two lowest eigenstates \cite{merzbacher}.

In an ideal scenario, where all types of excitations are suppressed, one expects the final 
condensate to be in the ground-state 
of the harmonic potential $V(x,T_{\rm m})$, with the corresponding energy being 
$E_{\rm id}^{\rm(gs)}=0.5$ and $E_{\rm int}^{\rm (gs)}=1.76$, for ideal and interacting 
gas respectively. In our simulations, however, although we have ensured adiabaticity, 
the final energy $E_{\rm id(int)}(T_{\rm m})$ of an ideal(interacting) gas may 
exceed the corresponding ground-state energy  $E_{\rm id(int)}^{\rm (gs)}$. 
In Fig. \ref{energylndscape}(c) we plot the excess energy in the system 
$U_{\rm id(int)}=E_{\rm id(int)}(T_{\rm m})-E_{\rm id(int)}^{\rm (gs)}$, as a function 
of the initial phase difference. 
In the presence of interactions, the excess energy is negligible for 
$\Delta\varphi_\text{in}<\pi/4$ and, to good accuracy, the system is in 
its ground state at the end of the adiabatic merging. For an ideal gas, 
the corresponding regime is much narrower as it pertains to very small 
phase differences $\Delta\varphi_\text{in} < \pi/8$. 
Moreover, in both cases we observe a rapid increase of the excess energy 
for increasing $\Delta\varphi_{\rm in}$, but for different reasons. 

In the case of a noninteracting gas, the problem under consideration reduces to the problem of a 
single particle in a time-varying double-well potential, because Eq. (\ref{tdgp}) reduces to the 
Schr\"odinger equation. Under the assumption of adiabatic merging, 
one may readily obtain analytic expressions for the occupation probabilities of the ground 
state ($p_0$) and the first excited state ($p_1$) of the final harmonic 
potential $V(x,T_{\rm m})$. The population imbalance is given by \cite{Zoltan}
\be p_0-p_1=\cos(\Delta\varphi_{\rm in}), 
\label{nonintimbalance}
\ee
and thus the excess energy (in units of $\omega$) reads 
\be
U_{\rm id}=\left [1-\cos(\Delta\varphi_{\rm in})\right ]/2.\label{nonintenergy}
\ee
This theoretical curve is also drawn in Fig. \ref{energylndscape}(c) and shows  very good 
agreement with our numerical results.  
Thus, in the case of an ideal gas, at the end of the merging the system is in a superposition of the 
ground state and the first excited state of the harmonic potential $V(x,T_{\rm m})$. 
In the extreme case of  $\Delta\varphi_{\rm in}=\pi$, the final state is basically the first excited 
state of the harmonic oscillator, i.e., the first asymmetric Hermite polynomial, which 
has a node at $x=0$.  Hence, the final density exhibits two distinct peaks, while it vanishes 
at the center of the trap. 

\begin{figure}
\includegraphics[width=8cm]{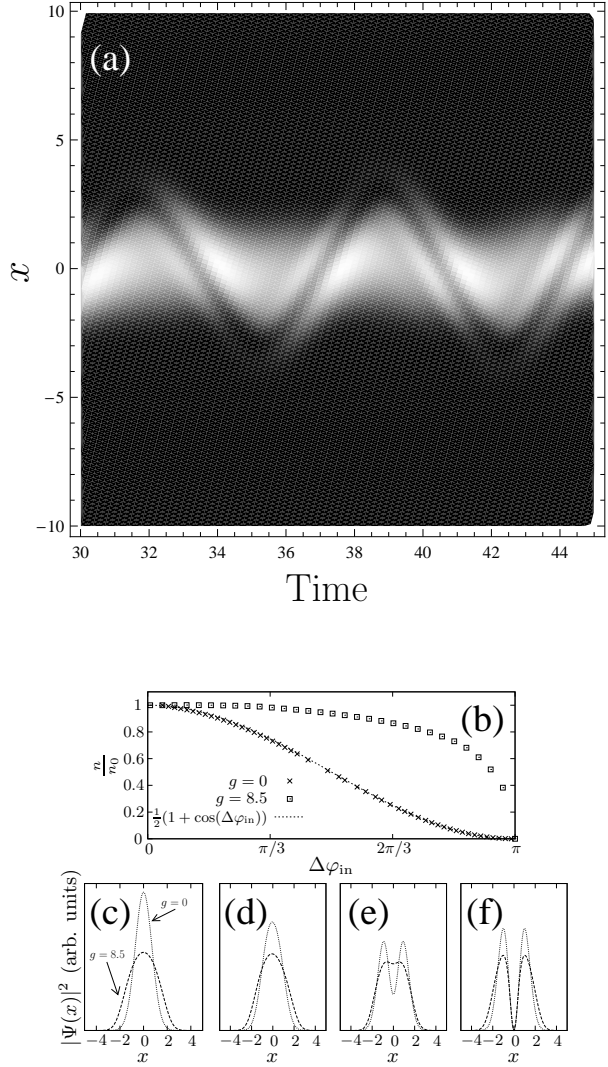}
\caption{
(a) Time-evolution of the atomic density distribution $|\psi(x,t)|^2$ in arbitrary units after the merging of two identical 
condensates with $\Delta\varphi_{\rm in}=3\pi/4$. Harmonic oscillator units are used for space and time.
(b) Relative depth of the dip in the final atomic density distribution at the center of the 
trap, for various values of $\Delta\varphi_{\rm in}$. (c-f) The entire distribution 
is depicted only for $\Delta\varphi_{\rm in}=0,\pi/3,2\pi/3$, and $\pi$.
}\label{soliton}
\end{figure}

In the case of an interacting gas, the excess energy observed in our simulations 
is due to a dark soliton which  is formed adiabatically during the merging process.  
The initial phase difference determines the depth of the soliton, and thus its dynamics. 
As before, merging is impossible for $\Delta\varphi_{\rm in}=\pi$, where we have the formation 
of a static black soliton and the 
final density in the single well exhibits two distinct peaks [see Fig. \ref{soliton}(f)]. 
For smaller phase differences, however, we have the formation of shallower solitons 
[Figs. \ref{soliton}(c-e)], 
which oscillate back and forth in the trap [see Fig. \ref{soliton}(a)]. 
This is in agreement with known results of soliton dynamics \cite{anglin}, and similar to the 
formation of vortices in the merging 
of three Bose-Einstein condensates with appropriate phase differences \cite{vortexmerge}. 

We have seen therefore that, for both ideal and interacting gases, the adiabatic merging 
results in a condensate whose density exhibits a dip. The origin of the dip, however, 
is fundamentally different in the two cases, and this fact is expected to be reflected in 
the dependence of the dip's depth on the initial phase difference. 
In Fig. \ref{soliton}(b) we plot the the dip's depth at $x=0$ (i.e., $n=|\Psi(0,t)|^2$ for $t>T_\text{m}$),
normalized to its value $n_0$ estimated for $\Delta\varphi_{\rm in}=0$, 
as a function of $\Delta\varphi_{\rm in}$. Using Eq. (\ref{nonintimbalance}), 
in the absence of interactions we obtain $n/n_0=[1+\cos(\Delta\varphi_{\rm in})]/2$. 
For an interacting gas, however, the ratio $n/n_0$ behaves differently for 
varying $\Delta\varphi_{\rm in}$.
In particular, we see that the dip remains rather shallow for $\Delta\varphi_{\rm in}\lesssim\pi/2$ 
and becomes deeper rather abruptly as we increase $\Delta\varphi_\text{in}$ further. 
In any case, our results show that one may get a rough estimate of the relative phase 
between two condensates by looking at the density profile after their adiabatic merging.

In closing, we would like to point out that Figs. \ref{energylndscape} and \ref{soliton}(b) 
can be extended to negative phase differences, taking the  mirror images of the corresponding curves for $\Delta\varphi_{\rm in}>0$, 
with respect to the vertical axis.   
This is due to the symmetry of the system under exchange of the two wells, which together with the global-phase invariance, 
imply that quantities that depend only on the density and the phase difference, are symmetric under the transformation 
$\Delta\varphi_{\rm in}\to -\Delta\varphi_{\rm in}$.
Indeed, the system is initially prepared in the state (\ref{gs_init2}) with real $\Phi_{\rm L(R)}(x)$. As discussed earlier, 
the dynamics of the system, from the point of view of densities and energies, do not change if we multiply the entire state  
by $e^{-{\rm i}\Delta\varphi_{\rm in}}$. Doing so, the initial condition reads 
$\Phi^\prime(x,t=0)=\Phi_{\rm L}(x)e^{-\text{i}\Delta\varphi_\text{in}}+\Phi_{\rm R}(x)$. 
Given that the symmetry of the system with respect to the two wells is preserved throughout the merging process,  
we can exchange the two initial wells obtaining   
$\Phi^{\prime\prime}(x,t=0)=\Phi_{\rm L}(x)+\Phi_{\rm R}(x)e^{-\text{i}\Delta\varphi_\text{in}}$, 
which differs from the initial condition (\ref{gs_init2}) by the sign of the phase difference.

\subsection{Center of Mass Motion}
\label{IIIB}
\begin{figure}
\includegraphics[width=8cm]{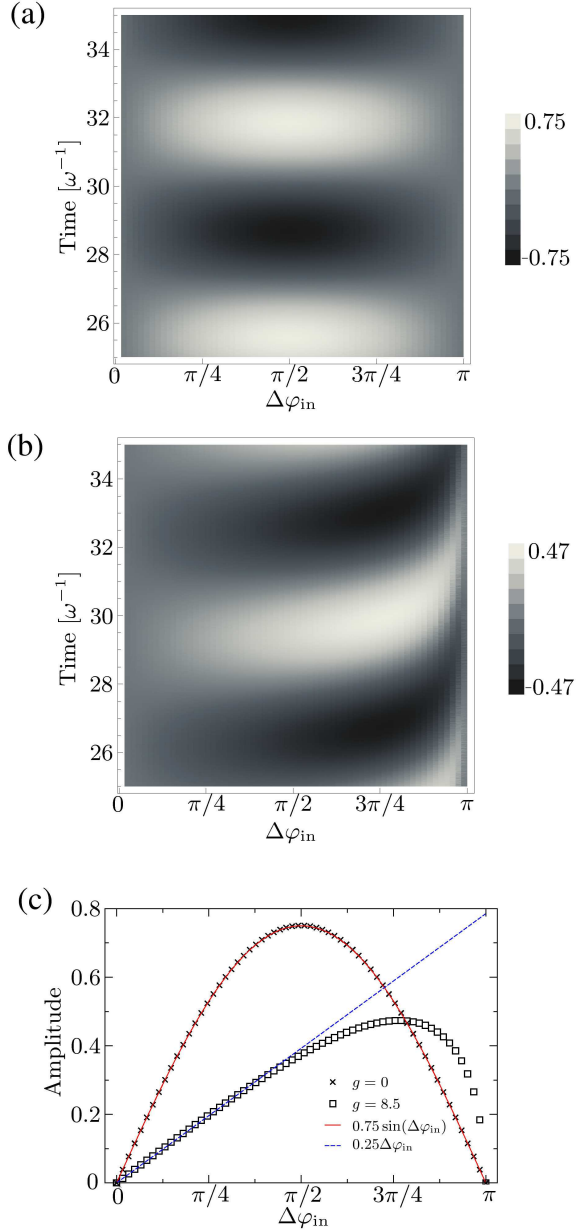}
\caption{(Color online) (a) Displacement of the center of mass in h.o. units with no interatomic interactions for $T_m=25\omega^{-1}$.
(b) Same as (a) but with interactions $g=8.5$. 
(c) Amplitude of the dipole oscillation in dependence of the initial phase difference $\Delta\varphi_\text{in}$.}\label{ampsfigure}
\end{figure}

To gain further insight into the nature of the excitations that occur 
during the merging, it is interesting to
investigate the dynamics of the center of mass motion \cite{review}
\begin{equation}
\bar{x}(t)=\int_{-\infty}^{\infty}{\rm d}x |\Psi(x,t)|^2 x.
\end{equation}
The evolution of $\bar{x}(t)$ as a function of time for various initial 
phase differences, is depicted in
Fig. \ref{ampsfigure}.
For all the choices of parameters, the behavior of  $\bar{x}(t)$ was 
found to be well approximated by
\be
\bar{x}(t)=C\sin[(t-T_{\rm m})+\theta],
\ee
with the amplitude $C$ and the phase $\theta$ depending on the initial 
phase difference
and the strength of interatomic interactions only.

As depicted in Fig. \ref{ampsfigure}(c), for $g=0$ the amplitude of the oscillations 
varies sinusoidally with
$\Delta\varphi_{\rm in}$, and is well approximated by 
\[C\simeq0.75\sin(\Delta\varphi_{\rm in}).\]
The presence of interactions deforms this symmetric behavior around 
$\pi/2$,
since $C$ attains its maximum for higher values of $\Delta\varphi_\text{in}$.
As far as the phase $\theta$ is concerned, for an ideal gas it depends on $T_m$ only, while for $g>0$ 
it also acquires a dependence on $\Delta\varphi_\text{in}$. 
According to our simulations, the excitations discussed in Sec. \ref{IIIA} (i.e., 
the presence of a soliton for $g\neq 0$,
or the population of the first excited state for $g=0$), cannot describe 
all aspects of the center of
mass motion. The above observations, as well as additional details not described here, show that besides the 
excitations discussed earlier, the
final condensate performs dipole oscillations with an amplitude 
depending on the initial phase difference.

\subsection{Final Phase} 
As we saw in the previous subsections,  for an interacting gas at the end of an adiabatic merging, 
we can distinguish between two types of phase-sensitive excitations in the final condensate, namely dipole 
and soliton-like excitations. 
In the one-dimensional model under consideration, such excitations 
do not decay, and thus the state of the final condensate cannot be expressed in the form of (\ref{gs}), 
or an easy modification thereof.

Recent experimental observations \cite{chikkatur, Jo07}, however, suggest a fast decay 
of phase-sensitive excitations in a three-dimensional merging setup, which results in an increase of the 
temperature on the order of $\hbar\omega/{\rm k}_{\rm B}\sim {\rm nK}$, where ${\rm k}_{\rm B}$ is the 
Boltzmann constant. This is in agreement with our estimates for the excess energy in the final condensate 
[see Fig. \ref{energylndscape}(c)]. This amount of energy can be dissipated by evaporative cooling, with a small 
loss in the number of atoms \cite{chikkatur, Jo07}. Most importantly, such a cooling mechanism is not expected 
to destroy the phase information carried by the final condensate; a very crucial issue for many 
applications (e.g., interferometry \cite{AtomInterf}, atom lasers \cite{cwAL,chikkatur}).

As an attempt to trace the final result of such a dissipation mechanism after the merging 
(i.e., for $t>T_{\rm m}$), 
we propagated Eq. (\ref{tdgp}) in imaginary time (i.e., replacing $\text{i}\partial_t$ by $\partial_t$).
This imaginary-time evolution, does not affect the complex phase of $\Psi(x,t)$, but only the 
density $|\Psi(x,t)|^2$ and appears as particle loss. Thus the function 
$\Psi(x,t)$ has to be continuously renormalized and ends up 
eventually in a stationary state of the form (\ref{gs}),  enabling us to 
extract the chemical potential and the phase $\varphi_{\rm f}$ of the final condensate.

\begin{figure}
\includegraphics[width=8.0cm]{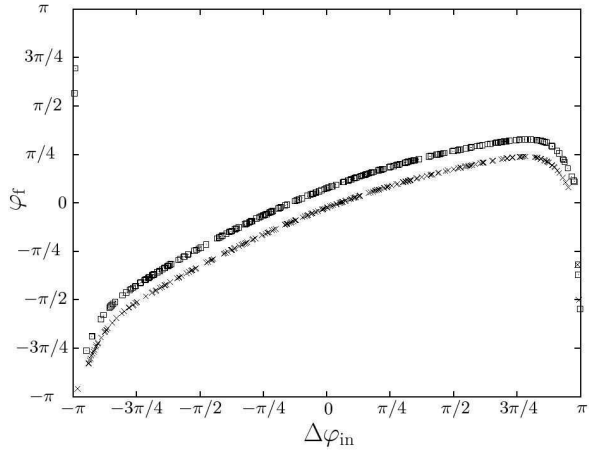}
\caption{(a) The phase of the final condensate $\varphi_{\rm f}$, as a function of the initial phase difference 
$\Delta\varphi_{\rm in}$ in the noninteracting case. Merging times are $T_{\rm m}=21.52\omega^{-1}$ (crosses) and $T_{\rm m}=26.52\omega^{-1}$ (squares). (b) Same as (a) but with interactions and merging times $T_{\rm m}=23 \omega^{-1}$ (crosses) and $T_{\rm m}=23.4 \omega^{-1}$ (squares). The line is Eq. (\ref{phaseformula}) applied to the upper curve.}\label{phase}
\end{figure}

According to our simulations, the final phase $\varphi_{\rm f}$ depends mainly on 
the merging time $T_{\rm m}$, the initial phase difference $\Delta\varphi_{\rm in}$, and the interaction strength $g$.  More interestingly, our simulations reveal the distinct roles of these parameters on the  
the phase of the final condensate.  
In Fig. \ref{phase}, we present results for ideal and interacting gases and for various merging times. These results pertain to the initial state (\ref{gs_init2}) where, without loss of generality, we have chosen as a reference phase $\varphi_{\rm L}=0$ and thus effectively 
$\Delta\varphi_{\rm in}=\varphi_{\rm R}$. 

\begin{figure}
\includegraphics[width=8.0cm]{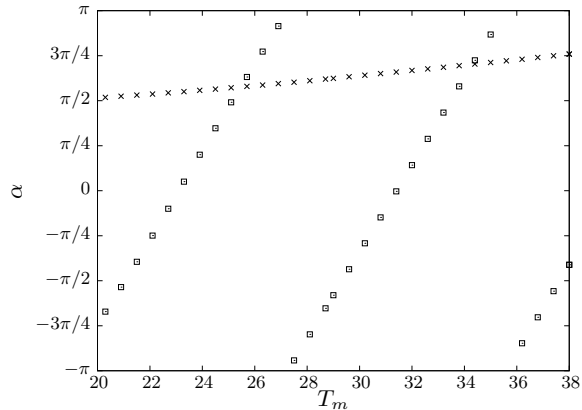}
\caption{
The function $\alpha$ in dependence of the merging time for interacting (squares) and non interacting (crosses) gases.
}\label{alphaplot} 
\end{figure}

In the absence of interactions, and for a given merging time, the final phase depends linearly on the initial phase difference $\Delta\varphi_{\rm in}$, with the corresponding 
slope being approximately equal to $0.5$ [see Fig. \ref{phase}(a)]. 
For increasing merging times, the entire curve shifts upwards linearly with $T_{\rm m}$, 
without any noticeable effect on its form. Hence we
have $\varphi_{\rm f}\approx \beta(0,T_{\rm m},0)+\Delta\varphi_{\rm in}/2$, where 
$\beta(g,T_{\rm m},\varphi_{\rm L})$ is a function of $g$, $T_{\rm m}$, and $\varphi_{\rm L}$. 
In view of the global-phase invariance discussed in Sec. \ref{SecIII}, the same results hold for 
arbitrary 
values of the reference phase $\varphi_{\rm L}$ with 
$\beta(g,T_{\rm m},\varphi_{\rm L})= \beta(g,T_{\rm m},0)+\varphi_{\rm L}$.  
Hence, the phase of the final condensate is well approximated by   
\begin{equation}
\varphi_{\rm f}=\alpha+\frac{\varphi_{\rm L}+\varphi_{\rm R}}{2}\label{phaseformula},
\end{equation}
where for the sake of simplicity we have set $\alpha(g,T_{\rm m})=\beta(g,T_{\rm m},0)$, 
with $\alpha$ being a linear function of the merging time (see Fig. \ref{alphaplot}) \footnote{It has to be pointed out that in Eq. (\ref{phaseformula}) all quantities are naturally defined only in the range 
$[-\pi,\pi)$.}.
Note that Eq. (\ref{phaseformula}) is invariant under the exchange of the two wells in 
agreement with the symmetries discussed in Sec. \ref{SecIII}.

For an interacting gas (i.e., for $g\neq0$), we have a similar behavior of the final phase as we vary all the relevant parameters. 
As depicted in Fig. \ref{phase}(b), for $|\Delta\varphi_{\rm in}|$ close to zero the final 
phase is well approximated by Eq. (\ref{phaseformula}). The parameter $\alpha(g,T_{\rm m})$ 
is a linearly increasing function of the merging time, with the slope being determined by the 
interaction strength $g$ (see Fig. \ref{alphaplot}).
For $\Delta\varphi_{\rm in}\approx\pm \pi$, however, we observe strong deviations of the 
final phase from Eq. (\ref{phaseformula}). Such deviations can be attributed to the nonlinearity, which 
gives rise to new phenomena that are not present in an ideal gas. For instance, as discussed in 
Sec. \ref{IIIA}, for $\Delta\varphi_{\rm in}\approx\pm \pi$, the merging of two independent condensates 
is impossible, due to the formation of a dark soliton.

A pertinent question is whether the behavior of $\alpha$ with respect to $T_{\rm m}$ can be 
explained solely by the movement of the traps. 
To answer this question, we have investigated the dynamics of a single atomic cloud in an adiabatically moving 
harmonic trap with 
\begin{equation}
V(x,t)=[x-s(t)\cdot l]^2/2\label{singlemove},
\end{equation}
where $s(t)$ is given by (\ref{s_t}). 
The movement takes place from $t=0$ to $t=T_{\rm m}$, and the condensate is prepared initially in the 
state $\Phi(x,0)$, which is the lowest-energy solution of the harmonic potential $V(x,0)=(x-l)^2/2$.

For a trap moving with a constant speed $v$, the wavefunction of the cloud in the laboratory frame, 
is given by 
\cite{LandauLifshitz} 
\begin{equation}
\Psi_{\rm lab}(x,t)=\psi(x-vt)e^{-{\rm i}(\mu +\frac{1}{2}v^2)t+{\rm i}vx}, 
\end{equation}
where $\psi(x)$ is the corresponding wavefunction in the moving frame. 
We see therefore, that the main effect of the movement is to increase the chemical potential 
by $v^2/2$ while creating a phase modulation (grating) determined by $vx$. Both of these terms 
will have influence on the phase of the cloud at the end of the movement.

In our model, the speed is not constant and can be defined as the rate at which the trap minimum  
changes in time, which according to Eq. (\ref{s_t}) is 
\begin{equation}
v(t)=\pm l \dot s(t)= \mp \frac{l\pi}{2T_{\rm m}}\sin\left(\frac{\pi t}{T_{\rm m}}\right)\label{v_T}.
\end{equation} 
Given the adiabatic nature of the motion, the phase shift acquired at $T_{\rm m}$ due to the 
term $v^2/2$ can be estimated as
\begin{equation}
\Delta\phi_{\rm move}\approx\frac{1}{2}\int_0^{T_{\rm m}}v(t)^2{\rm d}t, \label{moveshift}
\end{equation}
and using Eq. (\ref{v_T}) we obtain 
\begin{equation}
\Delta\phi_{\rm move}\approx \frac{\pi^2l^2}{16T_{\rm m}}\label{deltaphimove}. 
\end{equation}
For the range of parameters used throughout our simulations,  
$\Delta\phi_{\rm move}\sim 0.4$, while $\Delta\phi_{\rm move}\sim T_{\rm m}^{-1}$ as opposed to the 
linear increase of $\alpha$ with respect to $T_{\rm m}$.

\begin{figure}
\includegraphics[width=8.0cm]{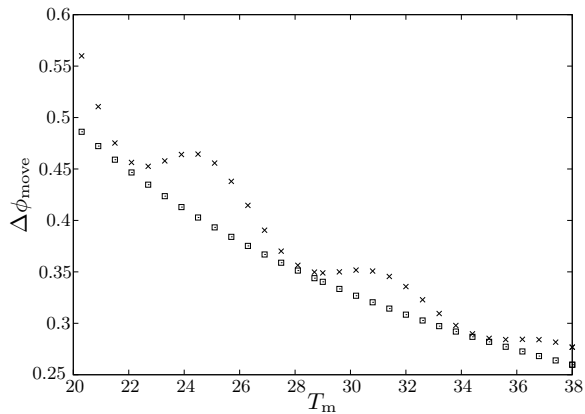}
\caption{
Phase shift experienced by a condensate trapped in a moving harmonic potential described by Eq. (\ref{singlemove}), as a function of the movement time for $l=4$. (crosses) Numerical solution obtained by propagating the GPE (squares). An estimation given by Eq. (\ref{deltaphimove}).
 }\label{singlemovefig}  
\end{figure}

The analytic treatment of the phase modulation $vx$ is far more complicated due to its temporal and  
spatial dependence. To investigate its role on the phase of a moving condensate, we have solved numerically 
the GPE for a condensate in the aforementioned adiabatically moving harmonic trap. 
The total phase shift that the condensate has acquired at the end of the movement can be read out in the way 
described in Sec. \ref{SecIII}, and is plotted in Fig. \ref{singlemovefig} 
for various values of $T_{\rm m}$, together with the estimation (\ref{deltaphimove}). 
Clearly, the main role of the product $vx$ is to add stepwise 
modifications on the power law of Eq. (\ref{deltaphimove}) and thus, in any case the linearly increasing 
behavior of $\alpha$ with respect to the merging time cannot be explained in the framework of 
moving independent condensates.  

In closing this section, we would like to note once more that the present theoretical framework cannot answer 
the question whether the phase of a condensate that was created by merging two independent condensates 
is completely randomized or depends on the initial relative phase. 
A random new phase would be a trivial case, renders the process equivalent to the creation of a new condensate from a thermal cloud. Assuming that this is not the case, we found a relation between the 
initial and the final condensates. 
Due to our numerical treatment, in particular due to the propagation in imaginary time, results concerning 
dependence on time have to be taken with a pinch of salt. Although our aim was to perform the merging adiabatically,  
it turned out that initial phase differences alone already lead to nontrivial excitations in the resulting condensate. 
Furthermore the extraction of the phase $\varphi_{\rm f}$ from the numerical results turned out to be very sensitive to 
even small excitations of the condensate. Resorting to imaginary time evolution,
our final result, given by Eq. (\ref{phaseformula}), separates the final phase of a condensate created by merging two independent condensates, in a global phase shift depending linearly on the merging time, and the arithmetic average of the two intial phases. The dependence on the experimentally well controllable merging time places the answer to questions concerning 
the final phase as well as the nature of the dissipation processes within reach of current experiments.

\section{Summary and Outlook}
We have investigated the adiabatic merging of two independent nearly identical 
Bose-Einstein condensates at zero temperature, within the framework of a one dimensional Gross-Pitaevskii 
equation. We have been able to answer some of the questions raised previously in the literature \cite{chikkatur, Jo07}, 
pertaining to the type of phase-sensitive excitations created during the merging, 
as well as the factors that determine the phase of the final condensate.

Our simulations show that the initial phase difference between the 
two condensates $\Delta\varphi_{\rm in}$, is a crucial parameter dominating an 
adiabatic merging process.  Irrespective of the strength of interatomic interactions,  
it may prevent merging altogether, with the final density distribution exhibiting two distinct peaks. 
For non-interacting gases, 
this is due to the fact that the final state is a superposition of the 
first two eigenstates of the Hamiltonian, the excited one having a node at $x=0$. 
In the interacting case, however, the separation is due to a soliton, i.e. a nonlinear phenomenon. 
For both cases, besides this type of excitations, we also have a dipole oscillation of the final
condensate. All of these excitations are phase-sensitive in the sense that their 
dynamics and characteristics are determined mainly by $\Delta\varphi_{\rm in}$. 
Our estimates for the excitation energy are in good agreement with recent experimental observations \cite{chikkatur, Jo07}.

Although in the one-dimensional model under consideration, phase-sensitive excitations have long 
life times, in realistic experimental setups they appear to decay quickly, increasing thus the 
temperature of the system. Removing the more energetic atoms by evaporative cooling, 
one may thus dissipate the excitation energy, without affecting the phase information carried 
by the final condensate. 
We simulated such a phase-information-preserving dissipation mechanism by propagating 
the Gross-Pitaevskii equation in imaginary time. In this way, we were able to analyze 
the phase of the final condensate, and show that for a given interaction strength it is determined mainly by the merging time,  
as well as the initial phase difference $\Delta\varphi_{\rm in}$. 
Moreover, we derived an analytic expression for this dependence. 
The verification of this formula as well as the underlying assumptions are technically 
within reach of current experiments. 
 
Finally, it should be emphasized that the above results are valid over a wide range of 
interatomic interaction strengths. There are still, however, many open issues pertaining 
to the merging of two condensates with different 
chemical potentials, and the decay of phase-sensitive excitations.  
These questions as well as the extension of the present one-dimensional model 
of merging to a full three-dimensional theory, are currently under investigation.

 \section{Acknowledgments} 
The work was supported by the EC RTN EMALI (contract No. MRTN-CT-2006-035369). GMN would like to acknowledge 
a very useful discussion with Prof. W. Ketterle during the Onassis Foundation Science Lectures in Physics 2007.

\end{document}